\documentclass{article}
\usepackage{spconf,amsmath,epsfig,amsfonts}
\usepackage{psfrag}
\usepackage{amsmath}
\usepackage{amssymb}
\usepackage{amsfonts}
\usepackage{color}
\usepackage{algorithm,algorithmic}

\def\ve#1{{\mathchoice{\mbox{\boldmath$\displaystyle #1$}}%
		      {\mbox{\boldmath$\textstyle #1$}}%
		      {\mbox{\boldmath$\scriptstyle #1$}}%
		      {\mbox{\boldmath$\scriptscriptstyle #1$}}}}

\newcommand{\argmin}{\mathop{\mathrm{argmin}}}
\def\PSNR{\mathrm{ PSNR}}
\def\punit{\, \mathrm}

\newcommand{\algorithmicinput}{\textbf{input:}}
\newcommand{\INPUT}{\item[\algorithmicinput]}
\newcommand{\algorithmicoutput}{\textbf{output:}}
\newcommand{\OUTPUT}{\item[\algorithmicoutput]}

\title{Content-Adaptive Motion Compensated Frequency Selective Extrapolation for Error Concealment in Video Communication}
\name{\vspace{-0.2cm}J\"urgen~Seiler and Andr\'e~Kaup\vspace{-0.1cm}}
\address{Chair of Multimedia Communications and Signal Processing, \\University of Erlangen-Nuremberg, Cauerstr. 7, 91058 Erlangen, Germany\\
{\{seiler, kaup\}@LNT.de}\vspace{-0.3cm}}

\begin{document}
\ninept
\maketitle

\setlength{\abovedisplayskip}{2mm plus1mm minus1mm}
\setlength{\abovedisplayshortskip}{1mm plus1mm minus1mm}
\setlength{\belowdisplayskip}{2mm plus1mm minus1mm}
\setlength{\belowdisplayshortskip}{1mm plus1mm minus1mm}


\begin{abstract} \label{abstract}
If digital video data is transmitted over unreliable channels such as the internet or wireless terminals, the risk of severe image distortion due to transmission errors is ubiquitous. To cope with this, error concealment can be applied on the distorted data at the receiver. In this contribution we propose a novel spatio-temporal error concealment algorithm, the Content-Adaptive Motion Compensated Frequency Selective Extrapolation. The algorithm operates in two stages, whereas at first the motion in a distorted sequence is estimated. After that, a model of the signal is generated for concealing the distortion. The novel algorithm is based on an already existent error concealment algorithm. But by adapting the model generation to the content of a sequence, the novel algorithm is able to exploit the remaining information, which is still available in the distorted sequence, more effectively compared to the original algorithm. In doing so, a visually noticeable gain of up to 0.51 dB PSNR compared to the underlying algorithm and more than 3 dB compared to other error concealment algorithms can be achieved.
\end{abstract}


\begin{keywords}
Error Concealment, Spatio-Temporal Extrapolation, Video Communication
\end{keywords}


\section{Introduction} \label{sec:introduction}

With an increased ability to compress digital video data and with still growing transport capacities, the transmission of video signals over wireless channels or the internet has become very popular. But, as such channels are not reliable, the risk of transmission errors or delayed transmission is always present. If hybrid video codecs such as the H.264/AVC \cite{ISO/IEC2003} are used, this will lead to severe distortions of the transmitted images. In order to cope with this, modern video codecs use two strategies \cite{Stockhammer2005}: error resilience to protect the bitstream against transmission errors and error concealment to conceal occurring distortions in the case that error resilience fails.

In the scope of this contribution, we will focus on error concealment techniques. In general, error concealment can be regarded as signal extrapolation task. In doing so, the video signal has to be extrapolated from correctly received regions into the region suffering from distortion. All existent error concealment algorithms can be divided into three groups: spatial, temporal, and spatio-temporal ones. The spatial error concealment algorithms only make use of correctly received signal parts which are located in the same frame as the distortion. In contrast to this, temporal ones as e.\ g.\ the Decoder Motion Vector Estimation (DMVE) from \cite{Zhang2000a} or the Extended Boundary Matching Algorithm (EBMA) from \cite{Lam1993} only use previously correctly received frames for reconstructing the distorted areas. The third group, the spatio-temporal algorithms exploit correctly received signal parts from the distorted frames as well as signal parts from previously received frames at the same time for concealing the distortion. The Improved Fading Scheme (IFS) from \cite{Hwang2008} or the Three-Dimensional Frequency Selective Extrapolation \mbox{(3D-FSE)} from \cite{Meisinger2007} are two examples out of this group. As the temporal correlation in a video signal usually is higher than the spatial correlation, in general, the quality of the concealment increases from spatial to temporal to spatio-temporal extrapolation \cite{Kung2006}. 

In \cite{Seiler2008a} we proposed another spatio-temporal error concealment algorithm, the Motion-Compensated Frequency Selective Extra\-polation (MC-FSE). This algorithm operates in two stages. At first the motion in the sequence around the distortion is analyzed. After that, the motion is revoked in order to align the data with respect to the distorted block, before a three-dimensional model of the signal is generated by utilizing an enhanced version of 3D-FSE. Although this algorithm has proven to be able to outperform many other temporal and spatio-temporal algorithms we have discovered that the extrapolation quality can be further improved by making the model generation adaptive to the sequence. In doing so, the information about the distorted signal, obtained from correctly received signal parts, is exploited more effectively and a higher concealment quality can be achieved. 

In the next section, the novel Content-Adaptive Motion Compensated Frequency Selective Extrapolation (CA-MC-FSE) is introduced in detail and the similarities and differences to MC-FSE are discussed. Afterwards, simulation results are given for demonstrating the improved extrapolation quality of the novel approach.


\section{Content-Adaptive Motion Compensated Frequency Selective Extrapolation}\label{sec:mcfse-wa}

\begin{figure*}
 	\centering
	\psfrag{x}[c][c][0.85]{$x$}
	\psfrag{y}[c][c][0.85]{$y$}
	\psfrag{t}[c][c][0.85]{$t$}
	\psfrag{m}[c][c][0.85]{$m$}
	\psfrag{n}[c][c][0.85]{$n$}
	\psfrag{p}[c][c][0.85]{$p$}
	\psfrag{A}[c][c][0.9]{$\mathcal{A}$}
	\psfrag{B}[c][c][0.9]{$\mathcal{B}$}
	\psfrag{P}[c][c][0.9]{$\mathcal{L}$}
	\psfrag{x0}[c][c][0.9]{$x_0$}
	\psfrag{y0}[c][c][0.9]{$y_0$}
	\psfrag{ttau}[c][c][0.9]{$t=\tau$}
	\psfrag{ttaum1}[c][c][0.9]{$t=\tau-1$}
	\psfrag{ttaum2}[c][c][0.9]{$t=\tau-2$}
	\psfrag{ttaup1}[c][c][0.9]{$t=\tau+1$}
	\psfrag{dtm2}[c][c][0.9]{\color{red} $\ve{d}^{\left(-2\right)}$}
	\psfrag{dtm1}[c][c][0.9]{\color{red} $\ve{d}^{\left(-1\right)}$}
	\psfrag{dtp1}[c][c][0.9]{\color{red} $\ve{d}^{\left(+1\right)}$}
	\psfrag{Video sequence}[l][l][1.0]{Video sequence}
	\psfrag{Motion compensated and}[l][l][1.0]{Motion compensated and}
	\psfrag{aligned extrapolation volume}[l][l][1.0]{aligned extrapolation volume}
	\psfrag{Loss area}[l][l][1.0]{Loss area $\mathcal{B}$}
	\psfrag{Support volume}[l][l][1.0]{Support volume $\mathcal{A}$}\vspace{-0.5cm}
 	\includegraphics[width=0.66\textwidth]{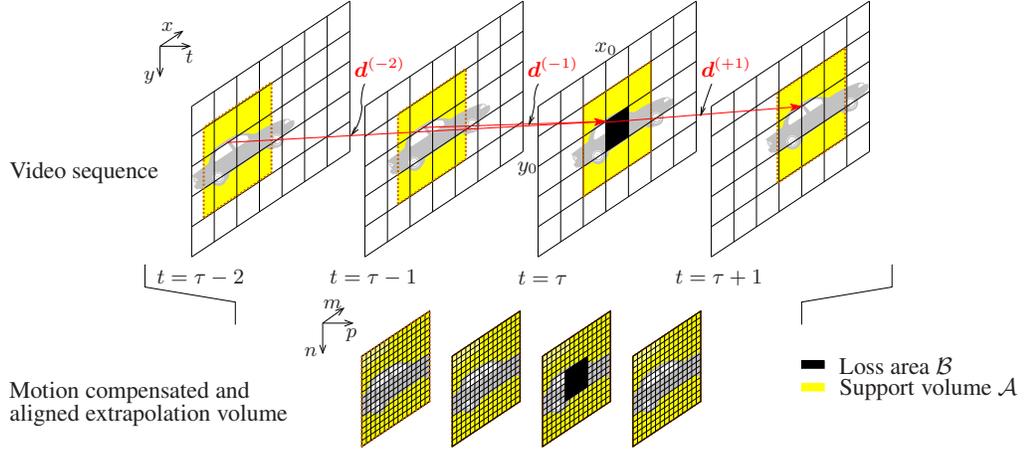}
 	\caption{Video sequence $v\left[x,y,t\right]$ and aligned extrapolation volume $\mathcal{L}$ (here: $N_\mathrm{p}=2$ and $N_\mathrm{f}=1$).}\vspace{-0.4cm}
 	\label{fig:extrapolation_area}
\end{figure*}

For outlining CA-MC-FSE, the scenario as presented in Fig.\ \ref{fig:extrapolation_area} top is regarded. Without loss of generality, an isolated block loss occurs in the frame at $t=\tau$ at position $\left(x_0,y_0\right)$ of video sequence $v\left[x,y,t\right]$. The sequence is depicted by spatial coordinates $x$ and $y$ and temporal coordinate $t$. For concealing the lost block, correctly received signal parts from the actual frame and, as far as available, from $N_\mathrm{p}$ previous and $N_\mathrm{f}$ following frames are exploited. The algorithm operates in two stages to conceal the loss. First, the motion in the sequence around the lost block is estimated. Afterwards, a data volume is cut out of the sequence which is aligned with respect to the estimated motion. Finally, a model of this aligned volume is generated by the already mentioned 3D-FSE from \cite{Meisinger2007}. The loss of isolated blocks is only considered for presentational reasons. In the case that other loss scenarios such as consecutive block losses occur, individual blocks are concealed successively. In doing so, already concealed blocks can be used as reference for unconcealed ones.

The first step, the motion estimation is necessary, since in most sequences, the content of a frame is at a different position in another frame. The relative displacement of a certain block from frame $t=\tau$ to frame $t=\tau+\kappa$ is depicted by the motion vector 
\begin{equation}
 \ve{d}^{\left(\kappa\right)} = \left(x_\mathrm{d}^{\left(\kappa\right)}; y_\mathrm{d}^{\left(\kappa\right)}\right)
\end{equation}
with horizontal displacement $x_\mathrm{d}^{\left(\kappa\right)}$ and vertical displacement $y_\mathrm{d}^{\left(\kappa\right)}$. As the content of the block at $\left(x_0,y_0\right)$ is lost, the content cannot be used for estimating the motion. To cope with this, an algorithm similar to the Decoder Motion Vector Estimation from \cite{Zhang2000a} is used to retrieve an estimate of the motion vector. In doing so, an area $\mathcal{M}$ of correctly received samples surrounding the lost block is regarded. Then, this area is compared to shifted versions of itself in the regarded reference frame. For this, for every possible candidate vector $\tilde{\ve{d}}^{\left(\kappa \right)}$ with a maximum displacement of $d_\mathrm{max}$, the error
\begin{equation}
E\left(\tilde{\ve{d}}^{\left(\kappa \right)}\right)  \hspace{-0mm}=\hspace{-1mm} \sqrt{\frac{1}{\left|\mathcal{M}\right|}\sum_{\forall \left(x,y\right) \in \mathcal{M}}\hspace{-2mm} \scriptstyle  \left(v\left[x,y,\tau\right] - v\left[x+\tilde{x}^{\left(\kappa \right)}_\mathrm{d}, y+\tilde{y}^{\left(\kappa \right)}_\mathrm{d}, \tau+\kappa\right] \right)^2}
\end{equation}
is determined for all pixels in $\mathcal{M}$. Then, from all candidate motion vectors, the one is selected that minimizes the error:
\begin{equation}
 \hat{\ve{d}}^{\left(\kappa \right)}  = \argmin_{\tilde{x}^{\left(\kappa \right)}_\mathrm{d}, \tilde{y}^{\left(\kappa \right)}_\mathrm{d} = -d_\mathrm{max}, \ldots, d_\mathrm{max}} E\left(\left(\tilde{x}^{\left(\kappa \right)}_\mathrm{d}; \tilde{y}^{\left(\kappa \right)}_\mathrm{d}\right)\right)
\end{equation}
Further, the estimation error 
\begin{equation}
\check{E}^{\left(\kappa \right)} = E\left( \hat{\ve{d}}^{\left(\kappa \right)}\right) 
\end{equation}
corresponding to the selected vector is stored for later usage. Performing these steps for all reference frames leads to a set of motion vectors $\hat{\ve{d}}^{\left(\kappa \right)}$ and corresponding estimation errors $\check{E}^{\left(\kappa \right)}$ for all $\kappa \in \left\{-N_\mathrm{p}, \ldots, N_\mathrm{f}\right\}\backslash 0$.

After the relative motion of the lost block to every regarded reference frame has been estimated, the quality of the estimation has to be evaluated, since the subsequent model generation is sensitive to a misalignment. As described in \cite{Seiler2008a}, two criteria are used for this. The first one is the absolute motion estimation quality, controlled by threshold $T_\mathrm{abs}$. The second criterion examines the homogeneity of the motion estimation and is controlled by threshold $T_\mathrm{rel}$.

The next step is to cut a three-dimensional data volume out of the sequence. The subsequent model generation by 3D-FSE generates a model of the signal suffering from distortion. For this, the so called extrapolation volume $\mathcal{L}$, centered by the loss, is taken from the video sequence. Volume $\mathcal{L}$ consists of the loss area, denoted by $\mathcal{B}$ and the support volume $\mathcal{A}$. The support volume on the one hand side contains correctly received signal parts surrounding the loss in the actual frame. On the other hand side, the corresponding areas in the reference frames, shifted according to the estimated motion are included. The volume is depicted by coordinates $m$, $n$ and $p$ and is of size $M\times N\times P$. The relation between video sequence $v\left[x,y,t\right]$ and extrapolation volume $\mathcal{L}$ is shown in Fig.\ \ref{fig:extrapolation_area} top and bottom. If the motion estimation is regarded to be unreliable, no alignment of the extrapolation volume is performed, and volume $\mathcal{L}$ consists of the samples at the same spatial positions in all reference frames.

After the extrapolation volume has been extracted from the video sequence, a model of the signal is generated using the Three-Dimensional Frequency Selective Extrapolation (3D-FSE) from \cite{Meisinger2007} which is enhanced by the fast orthogonality deficiency compensation from \cite{Seiler2008}. For extrapolating the signal from support volume $\mathcal{A}$ into loss area $\mathcal{B}$, 3D-FSE generates the parametric model
\begin{equation}
 g\left[m,n,p\right] = \sum_{\forall k \in \mathfrak{K}} c_k \varphi_k\left[m,n,p\right]
\end{equation}
which results from a weighted superposition of the mutually ortho\-gonal three-dimensional basis functions $\varphi_k\left[m,n,p\right]$. The weight of the individual basis functions is controlled by the expansion coefficients $c_k$ and the set $\mathfrak{K}$ contains the indices of all basis functions used for the model generation. The model is generated iteratively whereas in every iteration one basis function is selected and the corresponding weight is calculated. During the model generation, 3D-FSE uses the weighting function  $\rho\left[m,n,p\right]$ for controlling each pixel's impact on the model generation, depending on its position. With that, the influence each sample has on the model generation can be controlled. Thus, for example it is possible to give samples far away from the loss only a small weight and therewith small influence on the model generation whereas samples close to the loss get a high weight and therewith strong influence. To achieve this, the original 3D-FSE and MC-FSE give an exponentially decreasing weight to the samples with increasing distance. With $\hat{\rho}$ controlling the decay, the actual weighting results from
\begin{equation}
 \rho\left[m,n,p\right] = \hat{\rho}^{\sqrt{\left(m-\frac{M-1}{2}\right)^2 + \left(n-\frac{N-1}{2}\right)^2 + \left(p-N_\mathrm{p}\right)^2}}.
 \label{eq:old_weighting_function}
\end{equation}

But, by regarding common video sequences, it becomes obvious that the similarities between adjacent frames vary a lot and that the fixed weighting function used for MC-FSE can only be suboptimal. In the case that a reference frame is distinct to the frame containing the loss, the corresponding layer in extrapolation volume $\mathcal{L}$ should get a lower weight and vice versa.

The task is now, to determine the similarity between the frame suffering from the loss and the reference frames. To achieve this, the novel CA-MC-FSE reuses the motion estimation errors $\check{E}^{\left(\kappa\right)}, \forall \kappa \in \left\{-N_\mathrm{p}, \ldots, N_\mathrm{f}\right\}\backslash 0$ which have been calculated before. The motion estimation errors indicate, how well the area around the lost block can be located in a certain reference frame. By using the motion estimation errors, a shift of the content in the different regarded frames can be considered and does not affect the measure of the similarity.

As the exponentially decreasing weight used so far has already proven to lead to a decent extrapolation quality, the novel weighting function is based on the original one from (\ref{eq:old_weighting_function}). But, to account for the varying similarities between the individual frames, an additional weighting factor is added, controlling the weight in temporal direction. The factor $\Omega\left(\check{E}^{\left(p-N_\mathrm{p}\right)}\right)$ depends on the motion estimation error between the area around the loss and the corresponding area in frame $t=\tau+p-N_\mathrm{p}$. Thus, the function $\rho\left[m,n,p\right]$ which controls the influence of the correctly received samples, now is
\begin{equation}
 \rho \hspace{-0.5mm}\left[m,n,p\right]\hspace{-1mm} = \hspace{-1mm}\Omega\left(\hspace{-0.75mm}\check{E}^{\left(p-N_\mathrm{p}\right)} \hspace{-0.75mm}\right) \hat{\rho}^{\sqrt{\left(m-\frac{M\hspace{-0.5mm}-\hspace{-0.5mm}1}{2}\right)^2 \hspace{-0.5mm}+ \left(n-\frac{N\hspace{-0.5mm}-\hspace{-0.5mm}1}{2}\right)^2\hspace{-0.5mm} + \left(p-N_\mathrm{p}\right)^2}}.
\end{equation}
Next, the relationship between the motion estimation errors and the additional weighting of the exponentially decreasing weighting function has to be assigned. As a large motion estimation error should cause a small weight and a small motion estimation error should result in a large weight, we propose to use a linear decreasing correspondence between the motion estimation error and the additional weighting. With this, $\Omega\left(\check{E}^{\left(\kappa\right)}\right)$ is expressed by 
\begin{equation}
 \Omega\left(\check{E}^{\left(\kappa\right)}\right) = \left\{\begin{array}{ll} \Omega_\mathrm{max}\cdot\left(1-\frac{\check{E}^{\left(\kappa\right)}}{T_\mathrm{E}}\right) & \mbox{if }  \check{E}^{\left(\kappa\right)} < T_\mathrm{E} \\ 0 & \mbox{else }\end{array}\right.
\end{equation} 
for all $\kappa \in \left\{-N_\mathrm{p}, \ldots, N_\mathrm{f}\right\}\backslash 0$. Thereby, $\Omega_\mathrm{max}$ denotes the maximum value of the factor that can be achieved. The threshold $T_\mathrm{E}$ controls the maximum value of the motion estimation error. If this threshold is exceeded, the corresponding frame is regarded as completely distinct from the distorted one and is not included in the model generation step. Thus, for $\check{E}^{\left(\kappa\right)}$ varying between $0$ and $T_\mathrm{E}$, $\Omega\left(\check{E}^{\left(\kappa\right)}\right)$ varies between $\Omega_\mathrm{max}$ and $0$. 
 
Using such a simple correspondence between the weight and the motion estimation error has the advantage that the two values $\Omega_\mathrm{max}$ and $T_\mathrm{E}$ can be determined robustly from using training sequences. For this, for every candidate block, the best fitting weight is determined. Finally, a linear regression between the motion estimation errors and the best weights leads to $\Omega_\mathrm{max}$ and $T_\mathrm{E}$.

If consecutive block losses are considered instead of isolated block losses, the regions of the weighting function that correspond to yet unconcealed blocks are set to zero, as they cannot contribute anything to the model generation. Furthermore, the weight of the regions that correspond to already concealed blocks is attenuated by an additional factor $\delta$. This factor is smaller than one, in order to reduce the influence of already concealed regions on the model generation and therewith allay error propagation.

\begin{algorithm}
\caption{Error concealment by CA-MC-FSE}
\label{algo:mc-fse-wa}
\begin{algorithmic}
\INPUT distorted sequence
\FORALL {distorted blocks}
\STATE estimate motion vectors to all reference frames 
\IF {motion estimation is reliable}
\STATE align extrapolation volume according to motion vectors
\ELSE
\STATE use fixed extrapolation volume
\ENDIF
\STATE adapt weighting function to motion estimation errors
\STATE generate model by using 3D-FSE
\STATE cut out extrapolated block
\STATE replace lost block
\ENDFOR
\OUTPUT concealed sequence
\end{algorithmic}
\end{algorithm}

After the model generation has been finished, area $\mathcal{B}$ is cut out of the model and is used for concealing the loss. To give an overview of the individual steps needed for concealing a distorted sequence, CA-MC-FSE further is formulated as pseudo-code in \mbox{Algorithm \ref{algo:mc-fse-wa}}. For a detailed discussion of the model generation by 3D-FSE, please refer to \cite{Meisinger2007, Seiler2008a}.


\section{Simulation setup and results}\label{sec:results}

For evaluating the improvement of the novel CA-MC-FSE over \mbox{MC-FSE}, concealment of isolated and consecutive block losses is evaluated for various CIF-sequences in terms of PSNR between the reconstructed blocks and the undistorted ones. The distortions are lost blocks of size $16\times 16$ samples. The support volume for concealing the losses incorporates different numbers of $N_\mathrm{p}$ previous and $N_\mathrm{f}$ following frames. Further it consists of a frame of $16$ samples width around the lost blocks. The maximum search range for motion estimation is $d_\mathrm{max}=16$ and area $\mathcal{M}$ is $4$ samples wide. Furthermore, the thresholds for evaluating the motion estimation are $T_\mathrm{abs}=10$ and $T_\mathrm{rel}=3$. For generating the model, $800$ iterations are carried out, the decay factor $\hat{\rho}$ is set to $0.8$ and the orthogonality deficiency compensation factor is set to $\gamma=0.7$. If consecutive losses are considered, the additional weighting factor $\delta$ is set to $0.2$. The basis functions used, are the ones from the three-dimensional discrete Fourier transform. According to \cite{Meisinger2007}, this set is suited well for concealing errors in natural images as flat areas as well as edges and noise like areas can be reconstructed with high quality. This is due to the fact that the basis functions can be orientated in every direction and that shifts can be easily achieved by modifying the phase of a basis function. Additionally, by using this set, an efficient implementation working in the Fourier domain can be used. For transform into the Fourier domain, an FFT of size $64\times 64\times 16$ is used. The parameters listed so far are identical for MC-FSE and CA-MC-FSE. Fortunately, none of the parameters above is very critical or sequence dependent and they can be varied widely without heavily affecting the extrapolation quality.

As mentioned above, the two novel parameters $\Omega_\mathrm{max}$ and $T_\mathrm{E}$ can be derived from training sequences. For this, the uncoded sequences ``Discovery City'' and ``Discovery Orient'' are used and in every fifth frame between frame $5$ and $200$, $25$ blocks are concealed, every time with best possible weight. We use these two sequences, as they contain very different scenes, some with fast and some with slow motion, some which are very calm or others with illumination changes. Regarding the motion estimation errors for every block in combination with the best weight, the two parameters can be determined from a linear regression of $2000$ parameter pairs. The values resulting from the linear regression are $\Omega_\mathrm{max}=0.675$ and $T_\mathrm{E}=84.375$.

\begin{figure}
 	\centering
	\psfrag{Error pattern}[c][c][0.8]{Error pattern}
	\psfrag{MC-FSE}[c][c][0.8]{MC-FSE \cite{Seiler2008a}}
	\psfrag{MC-FSE-WA}[c][c][0.8]{CA-MC-FSE}\vspace{-0.4cm}
	\includegraphics[width=0.4\textwidth]{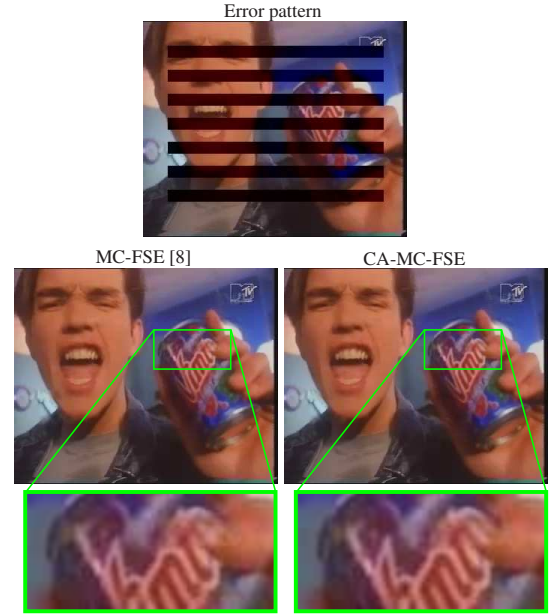}\vspace{-0.2cm}
 	\caption{Visual results for frame $10$ of sequence ``Vimto''.}
 	\label{fig:visual_results}
\end{figure}

In order to test the novel algorithm, it is implemented into the H.264/AVC reference software decoder JM14.0 \cite{JVT2008}. In doing so, artificial loss patterns are regarded. These patterns consist of isolated and consecutive blocks and correspond to the case that the checkerboard or the interleaved slice pattern is used. The artificial loss patterns are preferred over simulating packet losses as the latter produce unpredictable loss patterns which are not suited well for evaluating the actual extrapolation quality. The errors occur either in P-frames or in B-frames whereas for concealing errors in P-frames $N_\mathrm{p}=2$ previous and $N_\mathrm{f}=0$ following frames are used. If errors in B-frames are concealed, the support volume consists of $N_\mathrm{p}=2$ previous and $N_\mathrm{f}=1$ following frames. In addition to MC-FSE and CA-MC-FSE three other algorithms are also implemented for comparison: IFS from \cite{Hwang2008}, DMVE from \cite{Zhang2000a} and EBMA from \cite{Lam1993} on which the error concealment in the JM reference software is based. In Tab.\ \ref{tab:coded_conc} the results for concealing isolated blocks are listed for sequences ``Vimto'', ``Fastfood'' and ``Foreman''. Obviously, the two algorithms that make use of motion estimation and model generation are able to outperform DMVE in the mean by up to $3.65 \punit{dB}$, IFS by up to $2.92 \punit{dB}$ and EBMA by up to $4.89 \punit{dB}$. Comparing MC-FSE and the novel CA-MC-FSE one can further recognize that the concealment quality can be improved by up to $0.19\punit{dB}$ if P-frames are considered and up to $0.55\punit{dB}$ if the errors occur in B-frames.  Tab.\ \ref{tab:coded_conc_gain} further shows the mean gain of CA-MC-FSE over MC-FSE for the three sequences, if isolated or consecutive block losses occur either in P-frames or in B-frames. The results for consecutive block losses are comparable to the ones for isolated block losses whereas the gain is a little bit smaller. In summary, a mean gain of up to $0.51\punit{dB}$ can be achieved by using CA-MC-FSE. The larger improvement for B-frames can be explained by the circumstance that in this case more temporal information is incorporated due to the use of previous and following frames. With that, the adaptation of the weighting function to the sequence becomes more important and leads to a higher extrapolation quality. The enhanced extrapolation quality of \mbox{CA-MC-FSE} can also be recognized visually. Fig.\ \ref{fig:visual_results} shows frame $10$ from sequence ``Vimto'' as an example for concealment of consecutive block losses in B-frames. Regarding the whole frame, and especially the magnification of the text on the can, it becomes obvious that CA-MC-FSE produces less artifacts and provides an overall improved extrapolation quality.

\begin{table}
	\centering\vspace{-0.4cm}
	\fontsize{7}{9}\selectfont
		\begin{tabular}{l|c|c|c|c|c}
		& \hspace{-2mm}EBMA \cite{Lam1993}\hspace{-2mm} & \hspace{-2mm}DMVE\cite{Zhang2000a}\hspace{-2mm} & IFS \cite{Hwang2008}& \hspace{-2mm}MC-FSE \cite{Seiler2008a}\hspace{-2mm} &  \hspace{-1mm}CA-MC-FSE \hspace{-2mm}  \\ \hline
		\multicolumn{6}{l}{\hspace{-1.5mm}P-frames} \\ \hline
		\hspace{-1.5mm}``Vimto'' & $27.38 \punit{dB}$ & $25.78 \punit{dB}$ & $29.28 \punit{dB}$ & $30.42 \punit{dB}$ & $30.54 \punit{dB}$\\ \cline{1-6}
		\hspace{-1.5mm}``Fastfood''\hspace{-1.5mm} & $23.43 \punit{dB}$ & $23.02 \punit{dB}$ & $24.66 \punit{dB}$ & $25.96 \punit{dB}$ & $26.15 \punit{dB}$\\ \cline{1-6}
		\hspace{-1.5mm}``Foreman''\hspace{-1.5mm} & $28.46 \punit{dB}$ & $31.85 \punit{dB}$ & $30.45 \punit{dB}$ & $33.29 \punit{dB}$ & $33.33 \punit{dB}$\\ \hline
		\multicolumn{6}{l}{\hspace{-1.5mm}B-frames} \\ \hline
		\hspace{-1.5mm}``Vimto'' & $26.67 \punit{dB}$ & $26.02 \punit{dB}$ & $28.57 \punit{dB}$ & $31.22 \punit{dB}$ & $31.77 \punit{dB}$ \\ \cline{1-6}
		\hspace{-1.5mm}``Fastfood''\hspace{-1.5mm} & $23.91 \punit{dB}$ & $24.55 \punit{dB}$ & $25.61 \punit{dB}$ & $28.08 \punit{dB}$ & $28.60 \punit{dB}$ \\ \cline{1-6}
		\hspace{-1.5mm}``Foreman''\hspace{-1.5mm} & $29.80 \punit{dB}$ & $33.55 \punit{dB}$ & $32.11 \punit{dB}$ & $35.76 \punit{dB}$ & $36.23 \punit{dB}$ \\
		\end{tabular}\vspace{-0.1cm}
	\caption{Concealment results for isolated block losses.}
	\label{tab:coded_conc}
\end{table}

\begin{table}
	\centering
	\fontsize{7}{9}\selectfont
		\begin{tabular}{l|c|c}
		 & Isolated blocks & Consecutive Blocks \\\hline
		 P-frames & $0.12 \punit{dB}$ & $0.01 \punit{dB}$\\ \hline
		 B-frames & $0.51 \punit{dB}$ & $0.36 \punit{dB}$\\
		\end{tabular}\vspace{-0.1cm}
	\caption{Mean gain of CA-MC-FSE over MC-FSE for sequences ``Vimto'', ``Fastfood'', and ``Foreman''.}\vspace{-0.4cm}
	\label{tab:coded_conc_gain}
\end{table}


\section{Conclusion} \label{sec:conclusion}

In this contribution we introduced a novel spatio-temporal error concealment algorithm. The algorithm is based on an already very powerful extrapolation algorithm but is able to exploit the remaining information in a distorted sequence more effectively. This is achieved by adapting the model generation used during the concealment to the content of the distorted sequence. With that, the model generation can operate more precisely and a visually noticeable mean gain of up to $0.51 \punit{dB}$ in $\PSNR$ can be achieved compared to the original algorithm. Besides error concealment in video sequences, the proposed algorithm can be used for error concealment in multiview sequences as well. In this case, in addition to the motion in a sequence, the disparity between different views can be compensated before a four-dimensional model is generated.

\renewcommand{\baselinestretch}{0}
{\fontsize{9}{10}\selectfont

}

\begin{thebibliography}{10}
	\providecommand{\url}[1]{#1}
	\def\UrlFont{\rmfamily}
	\providecommand{\newblock}{\relax}
	\providecommand{\bibinfo}[2]{#2}
	\providecommand\BIBentrySTDinterwordspacing{\spaceskip=0pt\relax}
	\providecommand\BIBentryALTinterwordstretchfactor{4}
	\providecommand\BIBentryALTinterwordspacing{\spaceskip=\fontdimen2\font plus
		\BIBentryALTinterwordstretchfactor\fontdimen3\font minus
		\fontdimen4\font\relax}
	\providecommand\BIBforeignlanguage[2]{{%
			\expandafter\ifx\csname l@#1\endcsname\relax
			\typeout{** WARNING: IEEEtran.bst: No hyphenation pattern has been}%
			\typeout{** loaded for the language `#1'. Using the pattern for}%
			\typeout{** the default language instead.}%
			\else
			\language=\csname l@#1\endcsname
			\fi
			#2}}
	
	\bibitem{ISO/IEC2003}
	\mbox{Joint Video Team}~of {ISO/IEC MPEG} \&~{ITU-T VCEG}, ``Draft {ITU-T}
	recommendation and final draft international standard of joint video
	specification ({ITU-T} rec. {H}.264 | {ISO/IEC} 14496-10 {AVC}),''
	JVT-G050r1, Geneva, May 23-27 2003.
	
	\bibitem{Stockhammer2005}
	T.~Stockhammer and M.~M. Hannuksela, ``{H}.264/{AVC} video for wireless
	transmission,'' \emph{IEEE Wireless Communications}, vol.~12, no.~4, pp.
	6--13, Aug. 2005.
	
	\bibitem{Zhang2000a}
	J.~Zhang, J.~F. Arnold, and M.~F. Frater, ``A cell-loss concealment technique
	for {MPEG}-2 coded video,'' \emph{IEEE Trans. Circuits Syst. Video Technol.},
	vol.~10, no.~4, pp. 659--665, June 2000.
	
	\bibitem{Lam1993}
	W.-M. Lam, A.~R. Reibman, and B.~Liu, ``Recovery of lost or erroneously
	received motion vectors,'' in \emph{Proc. Int. Conf. on Acoustics, Speech,
		and Signal Processing (ICASSP)}, Minneapolis, USA, 27.-30. April 1993, pp.
	417--420.
	
	\bibitem{Hwang2008}
	M.-C. Hwang, J.-H. Kim, C.-S. Park, and S.-J. Ko, ``Improved fading scheme for
	spatio-temporal error concealment in video transmission,'' \emph{IEICE
		Transactions on Fundamentals of Electronics, Communications and Computer
		Sciences}, vol. E91-A, no.~3, pp. 740--748, March 2008.
	
	\bibitem{Meisinger2007}
	K.~Meisinger and A.~Kaup, ``Spatiotemporal selective extrapolation for 3-{D}
	signals and its applications in video communications,'' \emph{IEEE Trans.
		Image Process.}, vol.~16, no.~9, pp. 2348--2360, Sept. 2007.
	
	\bibitem{Kung2006}
	W.-Y. Kung, C.-S. Kim, and C.-C.~J. Kuo, ``Spatial and temporal error
	concealment techniques for video transmission over noisy channels,''
	\emph{IEEE Trans. Circuits Syst. Video Technol.}, vol.~16, no.~7, pp. 789--
	803, July 2006.
	
	\bibitem{Seiler2008a}
	J.~Seiler and A.~Kaup, ``Motion compensated frequency selective extrapolation
	for error concealment in video coding,'' in \emph{Proc. European Signal
		Processing Conference (EUSIPCO)}, Lausanne, Switzerland, 25.-29. Aug. 2008.
	
	\bibitem{Seiler2008}
	J.~Seiler and A.~Kaup, ``Fast orthogonality deficiency compensation for
	improved frequency selective image extrapolation,'' in \emph{Proc. Int. Conf.
		on Acoustics, Speech, and Signal Processing (ICASSP)}, Las Vegas, USA, 31.
	March - 4. April 2008, pp. 781--784.
	
	\bibitem{JVT2008}
	\mbox{Joint Video Team}, \emph{{H}.264/{AVC} reference software ({JM} 14.0)},
	http://iphome.hhi.de/suehring/tml/, 2008.
	
\end{thebibliography}
\end{document}